\documentclass[preprint, 12pt]{elsarticle}
\usepackage{lineno}
\usepackage{hyperref}
\hypersetup{colorlinks  = true}
\journal{Powder Technology}
\biboptions{authoryear}
\usepackage{amsmath}
\usepackage{caption}
\usepackage{subcaption}
\usepackage{graphicx}
\usepackage{upgreek}
\graphicspath{ {img/} }
\usepackage{geometry}
\usepackage{array}
\usepackage{upgreek} 

\usepackage{setspace}
 
\begin{document}

\begin{frontmatter}
	\title{Sedimentation effects on particle position and inertial deposition in $90^{\circ}$ circular bends}
	\author[label1]{Sara Vahaji}
	\author[label2]{Hien Nguyen}
	\author[label2]{Yidan Shang}
	\author[label2]{Kiao Inthavong \corref{cor1}}

	\cortext[cor1]{Corresponding Author: kiao.inthavong@rmit.edu.au}
	\address[label1]{School of Engineering, Deakin University, Geelong, Victoria 3217, Australia}
	\address[label2]{Mechanical \& Automotive Engineering, School of Engineering, RMIT University, Bundoora, Victoria 3083, Australia}

{\setstretch{1.0}	
\begin{abstract}
Laminar fluid-particle flows in bend geometries are present in many industrial, pharmaceutical, and biomedical applications. Particle deposition has been studied extensively; however, little attention has been paid to the effect of particle sedimentation on particle position and deposition in different pipe geometry combinations. This study presented a comprehensive analysis of sedimentation effects on particle flow behaviour in $90^{\circ}$ circular pipe bends of micron particles in laminar pipe flows. Pipe geometry combinations consisted of eight pipe diameters, nine bend radii, and 30 particle diameters in a range of 1 to 100 $\upmu$m. The results demonstrated the locations of particles that sedimented to the bottom half of the straight pipe section, and the particle positions upstream from the pipe bend entrance, which was no longer in a fully developed profile. These new locations represent the effects of gravity, pulling the particles down. While obtaining these positions can be found through CFD analysis, we proposed an analytical solution to predict the particle trajectory from different release locations that would help to identify the initial particle distribution at locations upstream to the bend, to obviate the need for long upstream straight pipe sections in the CFD analysis. 
\end{abstract}
}	
	
\end{frontmatter}

\section*{Introduction}
Dilute gas-particle flows in pipe bends are prevalent in many industries, including pharmaceutical, building ventilation, and oil and gas, where the particles are transported in pipes, ducts, or the human respiratory tract. In the human respiratory tract, pipe-like bends are observed at different locations: in the nostril to the main nasal passage, at the nasopharynx or the transition from the oral to oropharynx \citep{Yousefi2015,Storey-Bishoff2008, Inthavong2020, Li2013}. To improve the drug delivery efficiency, optimized designs exploiting gas-particle flow dynamics are needed; therefore, enhancing the knowledge of particle deposition in bends is critical. 

Particle deposition in pipes is influenced by flow rate, cross-sectional pipe area, and bend geometry. The pipe diameter could be as small as 0.001m (air sampling instruments) or as large as 1m (transporting fluids in industry), and the pipe bend can exhibit different curvatures. In many studies, the Stokes (St) number is reported as the influencing parameter on the particle deposition \citep{Shi2007}. Other investigations \citep{Grgic2004a} showed that Reynolds (Re) number also impacts the deposition process. These two parameters put together and recast as the inertial parameter, IP$ = d_p^2Q$ where $d$ is the aerodynamic particle diameter \citep{Inthavong2013}, and $Q$ is the flow rate, which is relevant for micron particle deposition that is susceptible to the inertial deposition mechanism. In addition to St and Re numbers, \cite{Dean1928} identified that the curvature ratio in bends has a direct impact on the particle deposition as well. The Dean (De) number was introduced as a function of Re number and the curvature ratio, where $\mathrm{De} = \mathrm{Re} / \sqrt{R_o}$. The curvature ratio is $R_o = R_b / R_p$ where $R_b$ is the bend radius, and $R_p$ is the pipe radius. However, all St, Re, and De numbers account only for impaction, and not the gravitational settlement (sedimentation). \cite{Nicolaou2018} introduced the Froude number, the ratio of the gravitational settling velocity to the fluid velocity, as the dimensionless number that accounts for both impaction and gravitation. 

To enhance the knowledge in the deposition process, many experimental  \citep{Johnston1973, Johnston1977,vahaji2015,Landahl1949,safari2018,Peters2004,Pui1987} and numerical \citep{Arsalanloo2017, Berrouk2007,Breuer2006,Cong2017,McFarland1997,vahaji2018,vahaji2019,Peng2016,Pilou2011, Tian2008, Tsai1990, Wilson2011} studies have been performed. The effects of particle size and flow rate on deposition was studied \citep{Cheng1999, Grgic2004}, as well as the effect of pipe bend geometry \citep{Grgic2004,Heenan2004,Nicolaou2013}, and different Reynolds numbers and curvature ratios \citep{Johnston1973,Johnston1977,Landahl1949, Pui1987}. 

Many investigations have also proposed analytical solutions for particle deposition \citep{Cheng1975,Cheng1981,Johnston1977,McFarland1997,Peng2016,Peters2004,Tsai1990,Wilson2011}; but these are limited to the range of parameters investigated and  may not translate well for general configurations. The author's previous work \citep{Inthavong2019} provided a correlation for micron particle deposition in pipe bends given as $\eta = 2 / \pi \tan^{-1} \left( 20.976 \mathrm{St}^{2.253} \right )$, where the Stokes number was defined as, $\mathrm{St} = \tau U_0 / R_p$; $\tau$ is the particle relaxtion time, $U_0$ is the mean velocity, and $R_p$ is the pipe radius. The results were based on a large data set that covered most common geometry arrangements and flow regimes in laminar flow. This included pipe diameters from 0.005m to 0.100m pipe diameters; bend radii from 0.006m to 0.070m; and Reynolds numbers of 400 to 2000 Reynolds numbers. 

In the presence of gravity, particle deposition in a pipe bend becomes sensitive to the upstream distance of particle release. \cite{PICH1973217} developed an analytical solution to predict the sedimentation deposition, but referred to it as the precipitation efficiency, in an elliptical and a circular channel as a function of a new parameter $\epsilon$. Similarly, a parameter, $k = (3/4) \mathrm{St} (gL/v_0^2)$ that could be used to characterize the deposition was proposed in \cite{finlay2008recent} based on an analytical solution for the gravitational settling of particles in a tubular geometry. 

This study builds upon the past work that has been achieved. For example, particle penetration in a laminar flow tube was investigated  \citep{Alonso2016}. Particle deposition in pipe bends have been studied extensively and for laminar flows some analytical solutions have been revealed, but under many assumptions that include an idealized flow. This includes \cite{Cheng1975, Cheng1981} which provided analytical solutions to the motion (of \cite{Mori1967}) and inertial deposition of particles in a bend, while \cite{davies1973}  demonstrated sedimentation in a straight pipe. More recently \cite{yu2021} reported diffusion and sedimentation effects in a straight, and \cite{Inthavong2019}'s results also showed scattered data due to sudden sedimentation from the different upstream release locations. 

Previous studies for deposition in a 90-deg pipe bends \cite{Li2013,guo2020, Inthavong2019, Breuer2006} used varying upstream particle positions where in real applications, heavy particles are influenced by sedimentation and the longer upstream distances will alter the particle positions as they enter the pipe bend section. This study aims to provide a technique to omit the upstream location by solving the sedimentation problem and comparing the solution with CFD predicted results. This will allow identification of particle positions due to sedimentation from the upstream straight pipe region, that end up affecting deposition efficiencies. This was  achieved by integrating an upstream horizontal straight pipe section to the pipe $90^{\circ}$ bend and considered the sedimentation effects on the overall deposition preceding the bend. CFD simulations of fluid-particle flow with  Re = 1000 for a combination of eight pipe diameters, nine bend curvatures, and thirty micron particles in the range of 1 to 100 $\upmu$m were performed to obtain an extensive data set for establishing deposition correlations.

The results are expected to serve as a reference point for future analysis of sedimentation and its effect on particle locations that influence deposition on a wall, while the analytical solution will be used as inlet boundary conditions for particle locations that consider sedimentation effects from upstream locations. This provides significant benefits for applications that exhibit long upstream locations.

\section{Method}
\subsection{Geometry Model}
Pipe geometries were created from combinations of pipe diameter $D$, with pipe bend radius $R_b$. For any combination, the geometry is physically constrained by $R_b > D/2$ (i.e. the bend radius must be greater than the pipe radius) otherwise it is physically impossible since the bend radius length $R_b$ from the horizontal and vertical pipe sections crosses over (see Fig 1a). Eight pipe diameters ranging from 0.005m to 0.100m combined with nine bend radii ranging from 0.006m to 0.070m were created (Table 1). A total of 48 cases were made after the constraint $R_b>D/2$.

{\renewcommand{\arraystretch}{2}
	\begin{table}[]
		\centering
		\hrule
		\begin{tabular}{p{5cm} l}
			\hline
			Pipe Diameter, $D$: & $5, 8, 10, 20, 40, 60, 80, 100$ \\
			Bend Radius, $R_b$: & $6, 8, 10, 15, 20, 30, 40, 55, 70$ \\
			and individual cases: & $D_1= 5.031$,  $R_{b1}=14.3355$  \\
			(for comparison against & $D_2=8.51$,  $R_{b2}=23.828$\\
			existing experiments)& $D_3=3.95$,  $R_{b3}=11.275$\\
		\end{tabular}
		\hrule 
		\vspace{0.5cm}
		\caption{Pipe geometry configuration (units in millimetres)}
		\label{tab:pipegeom}
	\end{table}
}

The inlet and outlet straight sections connected to the bend were made sufficiently long to avoid secondary flow influence from the bend curvature on the particle transport. Both \cite{Berrouk2007} and \cite{Breuer2006} applied a straight inlet section of 1D (one pipe diameter length) and a straight outlet section of 2D (two diameter lengths) to avoid disturbance on the flow in the bend. \cite{Roehrig2015} used 2.5D upstream and 1.5D downstream, while \cite{Zhang2012} used 3D lengths. \cite{Wilson2011} used 10D to represent the entrance length, but this was because the particles were uniformly distributed at the inlet. The entrance length was sufficiently long to allow the fluid and particles to fully develop over the entrance region. In this study, the straight pipe sections were set to 7D for the inlet and 3.5D  for the outlet which was sufficiently long to avoid an abrupt termination of the flow (Fig \ref{fig:geom}a). The influence of bend curvature on the upstream flow preceding the bend is expected to increases as the curvature ratio decreases to 1, i.e. $R_o \rightarrow 1$.  

The computational mesh used an octagonal block centred structured mesh, biased near the walls (Fig \ref{fig:geom}c). A mesh independence test was performed with three meshed models where the number of cells in each coordinate is given in Table \ref{tab:mesh}. The medium-mesh model was confirmed as independent through velocity profiles in pipe bends of $R_b = 2.8D$ (the velocity profile results are shown later in the Results section). The mesh density depended on the number of cells created along the pipe circumference, its radial distribution (biased towards the wall boundary), and distribution along its streamwise length. The total number of cells varied based on a constant distance between each cell along the bend curvature in the streamwise direction. Therefore, the largest pipe diameter with the largest bend radius produced the greatest number of cells. The number of cells in the radial and circumferential directions were constant. In the streamwise direction, the number of cells was equal to the straight inlet section (75 cells for medium mesh) plus the straight outlet section (45 cells for medium mesh) plus the number of cells in the bend which varied based on bend curvature defined as $(40 + R_b/D \times16)$. For example, this produced 176 cells for $R_b/D = 1.0$ and 205 cells for $R_b/D = 2.8$ for the Medium mesh settings.  

{\renewcommand{\arraystretch}{1.6}
	\begin{table}[h!]
		\centering
		\caption{Mesh configuration defined by the number of cells in the cross-section (radial, and circumferential directions) and along the streamwise direction.}
		\label{tab:mesh}
		\begin{tabular}{l c c c c c}
			\hline 
			\multicolumn{4}{c}{\textbf{number of cells}}&\multicolumn{2}{c}{\textbf{Example $R_b=0.028,  D=0.01$}}\\
			&radial&circ.&stream.&stream.&Total \\
			\hline
			Coarse&24&75&50 + (26 + $R_b/D \times 10$) + 30&134&241,200 \\
			Medium&36&112&75 + (40 + $R_b/D \times 16$) + 45&205&825,754 \\
			Fine&50&145&100 + (58 + $R_b/D \times 26$) + 60&291&2,108,300 \\
			\hline
		\end{tabular}
	\end{table}
}

\subsection{Fluid-particle modelling}
A laminar flow was used for the range of Reynolds numbers, and a fully developed profile was applied at the entrance,
\begin{equation}
	u(r)=2U_{ave}\left(1-\frac{r^2}{R^2} \right) 
	\label{eq:1}
\end{equation}
was used at the inlet for these studies, where $U_{ave}$ is the average velocity, $r$ is the radial distance from the pipe centre, and $R$ is the pipe radius. Reynolds number is kept at Re$ = 1000$ for all the cases, so for the smallest pipe diameter, $D_{min} = 0.005$m, maximum velocity of $U_{ave} = 2.92$m/s occurs. Similarly, the minimum velocity, $U_{ave} = 0.15$m/s, occurs in the largest pipe diameter, $D_{max} = 0.1$m. The fluid was treated as a continuous Eulerian phase and modelled using the CFD commercial code Ansys-Fluent v19v3. The discretisation schemes used were: second order for the pressure and momentum equation, and solving with the coupled approach. Particles were modelled using the Discrete Phase Model, which simultaneously tracks individual particles through the flow field. The particle phase was assumed dilute (< than 10\% volume fraction) such that the particle motion did not influence the fluid motion.  The particle either reached the pipe wall where it was assumed to deposit, or escaped through the computational domain via the outlet. \\
The particle equation of motion is given by
\begin{align}
	m_p\frac{d \vec{u_p}}{dt}=m_p\frac{\vec{u}-\vec{u_p}}{\tau_r}+m_p\frac{\vec{g}\left(\rho_p-\rho_g \right)}{\rho_p}+\vec{F}
\end{align}\\
where $m_p$ is the particle mass, $\vec{u}$ is the fluid phase velocity, $\vec{u_p}$ is the particle velocity, $\rho_g$ is the gas (air) density, $\rho_p$ is the density of the particle, $\vec{F}$ is an additional force, $m_p\frac{\vec{u}-\vec{u_p}}{\tau_r}$ is the drag force, $\tau_r$ is the droplet or particle relaxation time calculated by
\begin{align}
	\tau_r=\frac{\rho_p d^2_p}{18 \upmu}\frac{24}{C_d Re}
\end{align}\\
here, $\upmu$ is the molecular viscosity of the fluid, $d_p$ is the particle diameter, and $Re$ is the relative Reynolds number, which is defined as 
\begin{align}
	Re \equiv \frac{\rho d_p \vert \vec{u_p}-\vec{u} \vert}{\mu}
\end{align}\\
Also, the drag coefficient is taken from \cite{Morsi1972} defined by
\begin{align}
	C_d=a_1+\frac{a_2}{Re_p}+\frac{a_3}{Re_p}
\end{align}\\
where the $a_1$, $a_2$, and $a_3$ are empirical constants for smooth spherical particle over different ranges of particle Reynolds number. The particle Stokes number is
\begin{align}
	St=\tau \frac{U_o}{L_o}
\end{align}\\
${U_o}/{L_o}$ is the fluid time scale that is defined by the ratio of a characteristic velocity to a characteristic length, here it is taken as ${U_{ave}}/{R}$, where $R$ is the pipe radius (note that sometimes this length is taken as the pipe diameter).

Thirty particle diameters were considered, and for completeness and consistency, the entire range of diameters was used for all cases, despite some particles producing similar flow behavior. The diameters were: all (integers) between 1 to 20, 23, 25, 30, 35, 40, 45, 50, 60, 80, 100, giving thirty diameter sizes per pipe configuration. The deposition efficiency reported was for deposition in the bend section only, defined as
\begin{align}
	\eta = \frac{\text{number of particles depositing in bend}}{\text{number of particles entering the bend}}
\end{align}\\
thus, particle losses in the straight pipe section preceding the bend were excluded.

Particles were introduced at the inlet with an initial velocity matching the fully developed laminar flow profile in Eq \eqref{eq:1}. The total number of particles was tested for statistical independence by progressively increasing from 5000 particles to 50,000 and determining the deposition efficiency in one pipe condition. The final number for statistical independence was 30,000 particles. Deposition is based on parcel/particle mass depositing on the walls.

\subsection{Sedimentation}
Particle motion influenced by sedimentation was approximated in the mid-plane by adopting a force balance equation (Fig.\ref{fig:sediment}). Assuming a Stokes drag, and subjected to gravity, the particle motion in the vertical axis (y-axis) is given by:
\begin{align}
	\frac{dv_p(t)}{dt}=-C_1 v_p(t)-\beta g
\end{align}\\
where $v_p (t)=\frac{\left( dy(t)\right)}{dt}$. The solution for the vertical displacement is
\begin{equation}
	y_p(t)=y_0+\frac{\beta g \left( 1-e^{-C_1t}-C_1t \right)}{C_1^2}
	\label{eqn:yvel}
\end{equation}\\
where $y_0$ is the initial vertical position, $\beta= \left( 1-\frac{\rho_0}{\rho} \right)g$. 

The assumed Stokes drag law of $C_d = \frac{24}{{Re}}$ simplifies $C_1$ to the following
\begin{align*}
	C_1 =\frac{18\upmu}{\rho d_p^2}
\end{align*}
which is essentially the inverse of the particle relaxation time, e.g., $C_1 = 1/\tau$. For a particle moving near a wall, the drag force varies with the distance of the particle from the surface, and the drag force is adjusted to account for drag acting on a particle moving toward a wall under the creeping flow. This is given by \cite{Brenner1961} as:
\begin{align*}
	C_d =\frac{24}{\mathrm{Re}}A
\end{align*}
where
\begin{align*}
	A=1+\frac{d}{2h} 
\end{align*}
while for a particle moving parallel to the wall, \cite{Faxen1922}'s equation gives 
\begin{align*}
	A=\left(1- \frac{9}{16} \left(\frac{d}{2h}\right) + \frac{1}{8} \left(\frac{d}{2h} \right) ^3 - \frac{45}{256} \left(\frac{d}{2h}\right)^4- \frac{1}{16} \left(\frac{d}{2h}\right)^5  \right)^{-1}
\end{align*}

This suggests the particle relaxation time also needs adjustment and this value is tuned against the CFD results, and for now is simply given as 
\begin{align}
	C_1 =\frac{18\upmu}{\rho d^2}A_1
	\label{eqn:A1}
\end{align}

Particle motion in the streamwise direction (see \ref{fig:sediment}) was determined by
\begin{align}
	\frac{dw_p(t)}{dt}=C_1 \left[ w_f(t)-w(t) \right]  
	\label{eqn:streamA1}
\end{align}
The solution for the horizontal displacement is therefore,
\begin{align}
	z_p(t) = & C_6-\frac{C_2C_4^2e^{-2C_1t}}{2C_1}-C_2\left( -1+C_3^2+4C_3C_4+5C_4^2\right) t\\ \nonumber
	&  +C_1C_2C_4\left( C_3+2C_4\right) t^2-\frac{1}{3}C_1^2C_2C_4^2t^3\\ \nonumber
	&  +\frac{e^{-C_1t}\left\lbrace -C_5+C_2C_4\left[ C_1^2C_4t^2-2C_3\left( 1+C_1t\right) \right] \right\rbrace }{C_1}
\end{align}
where the constants $C_1$ to $C_6$ are:
\begin{align*}
	C_1 &= \frac{18\upmu}{\rho d^2}A_1\\ 
	C_2 &= 2w_A\\ 
	C_3 &= \frac{y_0}{R}\\ 
	C_4 &= \frac{\beta g}{R C_1^2}\\ 
	C_5 &= 2w_A \left( 1-\frac{y_0^2}{R^2} \right) +C_2 \left( -1+C_3^2+4C_3C_4+4C_4^2 \right)\\ 
	C_6 &= \frac{C_2C_4 \left( 4C_3+C_4 \right)+2C_5}{2C_1}
\end{align*}

\section{Results}
\subsection{Velocity Profiles}
Fig \ref{fig:valid} presents the radial profiles of velocity magnitude along lines A-A' (Fig \ref{fig:valid}a), and B-B' (Fig \ref{fig:valid}b) taken at $\theta=45^\circ$ and $\theta=90^\circ$. In this figure, our results for $R_o=5.6$ are compared to numerical results – based on a Finite Volume computational model – from Nicolaou and Zaki (2016). Fig \ref{fig:valid}a shows that at $\theta=45^\circ$, due to the inertial forces the velocity profile is skewed towards the outer wall. At the bend, secondary flow forms and continues to the bend exit (at $\theta=90^\circ$) where another peak with less magnitude is formed near the inner wall. Line B-B’ (Fig \ref{fig:valid}b) presents a symmetrical velocity profile for both $\theta=45^\circ$ and $\theta=90^\circ$ locations. Due to the vortices that are shaped following the secondary flow, the velocity profile slightly changes at $\theta=90^\circ$, which is compatible with the observations at line A-A’.

Fig \ref{fig:streams} shows the flow development across the pipe bend for $R_o=2.0$ (sharp bend) and $R_o=14.0$ (long bend). The effect of secondary flow on the streamlines and vortex structures are depicted, where the streamlines split in the middle and diverge to the sides to produce the vortex structures. The effect of secondary flow is evident in the streamlines. In the sharp bend, disturbances in the fully developed laminar velocity profile occur well before the inlet of the bend section. In contrast, the disturbances in the long bend are contained within the pipe bend itself. This suggests a minimum upstream distance should be considered in pipe bends that have high Dean numbers. 

To better understand the flow field for different curvature ratios, velocity contours were extracted in the pipe bend at angles $\theta=0^\circ$, $22.5^\circ$, $45^\circ$, $67.5^\circ$, and $90^\circ$ shown in Fig \ref{fig:velCont} pipe bend ratios of $R_o=$1.2, 1.6, 2.0, 3.0, 4.0, 6.0, 8.0, 11.0, and 14.0. Since the inertia is constant (constant Re number), the differences in the cases are caused only by centrifugal forces by changing the curvature ratio.

The cross-section at the bend inlet plane, $\theta=0^\circ$, shows the flow field is affected by sharper pipe bends (e.g. when $R_o$ is small, $\leq 2$) depicted in the pipe bends with a lower curvature ratio. The fully developed profile (where the peak flow is at the centre of the pipe) is altered to having lower velocities at the inner wall and peak velocity on the outer wall (top of each cross-section). The secondary flow effects are diminished with increasing curvature ratio. In sharp bends, the short pipe circumference prohibits the development of Dean vortices where become are not present until the bend exit at $\theta=90^\circ$. With the longer bends, the centrifugal forces have sufficient distance to transfer the momentum into the secondary flow, leading to Dean vortices occurring progressively earlier (with increasing $R_o$) in the pipe bends, i.e. occurring at cross-sections closer to the pipe bend inlet.

\subsection{Particle Deposition Efficiency}
Deposition efficiency for three pipe combinations of $R_b$, and $D$(5.03 mm, 3.95 mm, 8.51 mm) were used to obtain $R_o = 5.5, 5.6$, and 5.7 following the experimental setup of \cite{Pui1987}. The deposition efficiency in the pipe bend for Re=1000 was compared against studies in literature, with gravity being considered (Fig \ref{fig:depvalid}a) and without gravity (Fig \ref{fig:depvalid}b). There was a good comparison with the literature, especially for higher Stokes numbers: considering gravity and with \cite{Pui1987}, and neglecting gravity with \cite{Breuer2006}.

The influence of sedimentation can be depicted by varying the particle release locations (from 1D, 3D, and 5D downstream of the inlet) and tracking the particle positions in planes at sequential pipe diameter distances downstream. Two distinct particle sizes, $4\upmu$m and $30\upmu$m representing both low and high inertial properties and light and heavy particles were investigated. Fig \ref{fig:pdep11} shows the particle positions inside the pipe and wall deposition, with each particle coloured by velocity in a pipe with a diameter of 0.01m and a radius of 0.1m

Fig \ref{fig:pdep11}a demonstrates the behaviour of 4 $\upmu$m particles released from different upstream locations (at 1D, 3D, and 5D). The cross-section planes in the pipe inlet section show the particles maintain an entrained state within the transporting air. Particles at the top of each plane remain close to the top pipe surface, suggesting negligible sedimentation effects. Particle deposition on the pipe surface is sparse and found only in the bottom half of the pipe inlet section and the inner pipe bend section. In the pipe bend section, the particles continue to behave like a gas tracer moving with the air and tracing out the Dean vortex patterns. The outlet plane depicts a high number of particles escaping for all release locations. This suggests that low inertial particles are insensitive to sedimentation effects, and therefore the upstream release location are negligible for the distances tested up to 7D in length.

Conversely, the behaviour of 30 $\upmu$m particles is sensitive to the upstream release location (Fig \ref{fig:pdep11}b) due to the influence of sedimentation. The cross-section planes in the pipe inlet section show the particles have descended and the top-most particles of each plane are clearly below the top pipe surface. Particle deposition on the bottom surface of the pipe inlet section is more concentrated. Deposition at the pipe bend section occurs on the outer pipe wall suggesting inertial impaction. In the pipe bend section, the number of particles travelling in the pipe is reduced. The outlet plane depicts a very sparse pattern suggesting a small number of particles escaping for all release locations. This indicates that 30 $\upmu$m particles are sensitive to sedimentation, and the upstream release location has a significant influence on the particle distribution downstream.

Fig \ref{fig:pdep14} demonstrates the effects of greater Dean number flows where the the pipe bend is shortened by a factor of four (e.g., from $R_b =0.04$m to $R_b =0.01$m). In both cases, the effects of sedimentation for both 4 and  30$\upmu$m are the same.  For 30$\upmu$m particles, significant impaction occurs on the outer pipe bend wall, reducing the number of particles moving through the outlet pipe sections. As Dean number is increased, we expect to see an increased intensity of the secondary flow, and this is depicted by the 4$\upmu$m particles moving through the pipe bend and the outlet plane. The intense secondary flow from the pipe bend penetrates the upstream pipe bend inlet locations (at the 7D location) where there is a small crescent of particles located above the main circular plane of particles.

The effect of particle release location on the deposition efficiency of the bend wall is given in Fig \ref{fig:depcurves}a, where the results in each panel show the pipe diameter change from 0.01m to 0.1m, and profiles are coloured by the particle release location for all the bend curvature ratios. The results show that an increase in the pipe diameter, regardless of the curvature ratio, the sedimentation effects increases. The deposition efficiency curve collapses due to fewer particles depositing on the wall bend, even for the particles that are released close to the bend (at location 6D). For particles released closer to the pipe bend inlet, deposition was due to inertial impaction on the outer wall.

The effect of bend curvature ratio on the deposition efficiency of the bend wall is given in Fig \ref{fig:depcurves}b where the colours represent different curvature ratios. An increase in the curvature ratio produces an increase in deposition efficiency. This suggests that in addition to the inertial impaction on the outer wall, the particles will deposit on the inner wall due to momentum loss in the higher curvature ratio pipes (e.g., longer pipe bends). 

\subsection{Particle Deposition and Sedimentation Analysis}
To obtain an overall correlation, the data for all particle diameters and all cases were plotted (Fig \ref{fig:correlations}). For pipe bend deposition efficiency the particle Stokes number was used as the normalising parameter, and for sedimentation, it was $\epsilon$ from \cite{PICH1973217}. Data points are semi-transparent and coloured by particle diameter, while the solid curve fit line was taken from \cite{Inthavong2019} which is defined as $2/\pi \tan^{-1} \left(a St^b \right)$. The curve fit captures the overall inertial impaction onto the pipe bend section. However, there is some scatter of the data, which is primarily limited to larger particles (coloured in red). The effects of sedimentation were evident in larger pipe diameters for large particles (shown in Fig \ref{fig:depcurves}) where the deposition and particle distribution in cross-section planes were unique based on upstream particle release locations. Therefore, the curve fit line from \cite{Inthavong2019} to describe deposition efficiency (Fig \ref{fig:correlations}a) has limitations in its ability to capture the effects of sedimentation on larger particles, i.e. greater than 20$\upmu$m, and that the upstream locations should be considered when the gravitational force is involved.

Sedimentation in the straight pipe inlet section (before the bend) was plotted for all particles in all pipes. The analytical curve for Fig \ref{fig:correlations}b was from \cite{PICH1973217}, defined as 
\begin{equation}
	2/\pi \left( 2\epsilon \sqrt{1-\epsilon^{2/3}}-\epsilon^{1/3} \sqrt{1-\epsilon^{2/3}} + \arcsin{\epsilon^{1/3}} \right)
\end{equation}

where $\epsilon= \frac{3}{4} \frac{V_s}{D} \frac{L}{U}$ is the dimensionless parameter, $V_s$ is the stationary sedimentation velocity of particles, and L is the length of the inlet section. The deposition efficiency is dominated by the larger particles, while particles smaller than 20$\upmu$ having less than 0.15 deposition efficiency. The analytical curve fit equation predicts the particle sedimentation on the upper range and in most cases, overpredicts the simulation results. 

The particle trajectories for 4$\upmu$m, 15$\upmu$m, and 50$\upmu$m in three pipe geometries (short pipe bend, long pipe bend, large pipe diameter) and released from four heights ($R$, $R/2$, $0$, and -$R/2$) in the pipe are presented in Fig \ref{fig:sedimentTraj}. The CFD results were plotted in black lines. A comparison was made with the analytical solution given in Eqn. \ref{eqn:yvel} where $A_1$ in the vertical motion (Eqn.\ref{eqn:A1}) was 2/3, while in the streamwise motion (Eqn.\ref{eqn:A1}) was 1. The trajectories from Eqn. \ref{eqn:yvel} were coloured red, blue, green and pink, and overlayed onto the CFD results which demonstrated very good matching of results. Small deviations were found at the end of the straight pipe section at a distance 7D from the inlet where secondary flow effects influenced the flow field. 

The trajectories demonstrate the amount of vertical drop due to sedimentation over the straight pipe section. For the larger pipe diameter, the average flow velocity was  reduced to maintain the same Reynolds number. Therefore, the sedimentation is substantially greater, particularly for 15$\upmu$m and 50$\upmu$m particles. For the pipe diameters of $D=0.01m$ 50$\upmu$m particles released from the top of the pipe, at \textit{R}, drop much faster than the particles released at \textit{R/2}  because the flow field exhibits a fully developed laminar velocity profile. The trajectories indicate the top-most bounded particle positions found in the downstream planes (visualised in Fig. \ref{fig:pdep11}b and \ref{fig:pdep14}b)  after sedimentation. As a result, the analytical equation could determine the particle positions at a location downstream from any upstream location that could be very long, e.g., 10D, 20D, or 40D. This can allow a standard straight pipe section length to be used, typically 3D in length with a fully developed profile and the initial particle positions due to upstream sedimentation is defined by Eqn. \ref{eqn:yvel} and Eqn. \ref{eqn:streamA1}.

\section{Conclusion}
In this study, the effect of sedimentation in laminar fluid-particle flows in pipe bend combinations is investigated. Particles ranging from 1 to 100$\upmu$m  were released at cross-sections planed with a fully developed laminar profile. Different release cross-section locations from the inlet were evaluated to analyse particle deposition on the pipe surface. The findings are especially important for the applications where long upstream locations are present. A detailed study identifying the particle positions affected by sedimentation from upstream locations will reduce the computational cost by removing the need for CFD modelling of the upstream pipe sections. The analytical solution can be used to determine the initial particle positions in the straight pipe section, at different upstream locations from the pipe bend entrance.  

The results showed that small particles (4 $\upmu$m) travelled with the flow field unless they were located in low-velocity regions (such as the near wall, or in the recirculation zones). In contrast, larger particles were influenced by inertial impaction and sedimentation. Releasing particles close to the bend inlet were subjected to secondary flow effects in high Dean numbered flows, suggesting that two to three pipe diameters of a straight section should be used. In pipes with larger diameters, the average flow velocity was reduced to maintain a consistent Reynolds number, and this led to increased particle sedimentation. This was also found in bends with higher curvatures, where the particles deposited on the inner wall of the bend due to momentum loss. 

Visualisation of the particle positions through the pipe cross-sections demonstrated the influence of sedimentation on the larger particles, while the smaller particles traced out the flow field effectively, particularly the Dean vortices after the pipe bend. An analytical solution to particle motion in a straight pipe was proposed which predicted the same particle trajectories found in the CFD simulations. The equation provides a technique to predict particle positions downstream from the inlet of very long straight pipes and therefore, gives computational efficiencies by avoiding the CFD modelling of an entire straight section where particle deposition in pipe bends are of interest.

\section{Acknowledgements}
The authors gratefully acknowledge the financial support provided by Deakin University startup funds.

\small
{\setstretch{1.0}
\bibliography{sedimentation}

\normalsize
\newgeometry{margin=2.0cm}
\clearpage
\begin{figure}
	\centering
	\begin{subfigure}[b]{0.4\textwidth}
		\includegraphics[width=\textwidth]{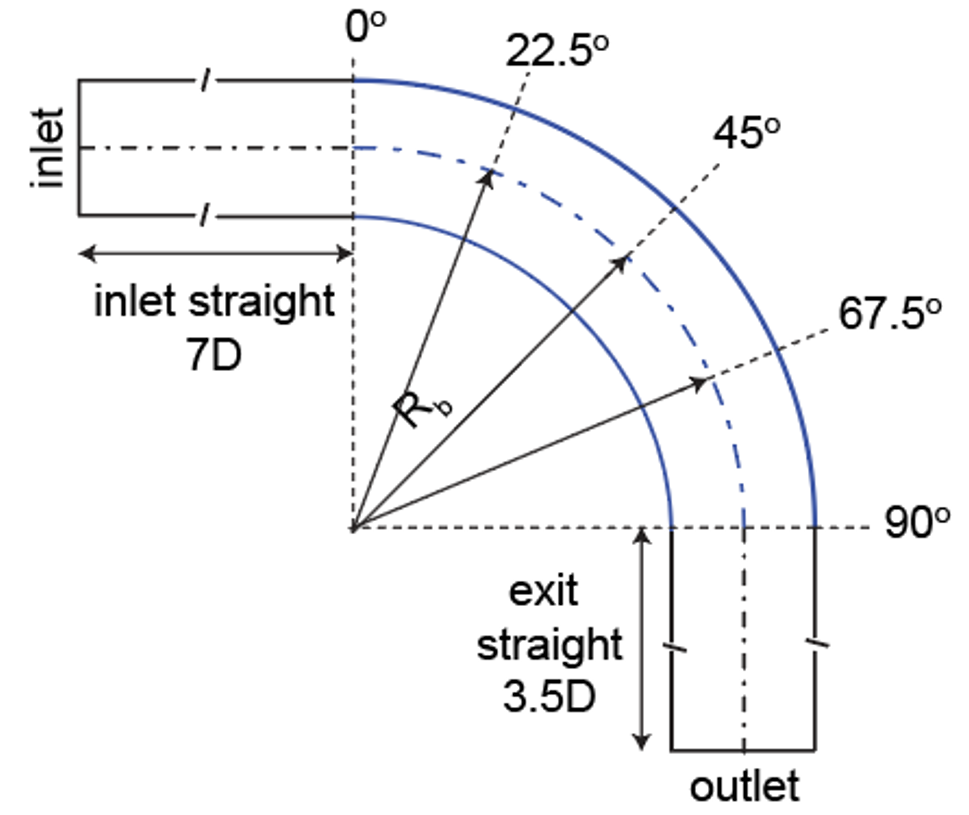}
		\caption{}
		\vspace*{4mm}
	\end{subfigure}
	~
	\begin{subfigure}[b]{0.49\textwidth}
		\includegraphics[width=\textwidth]{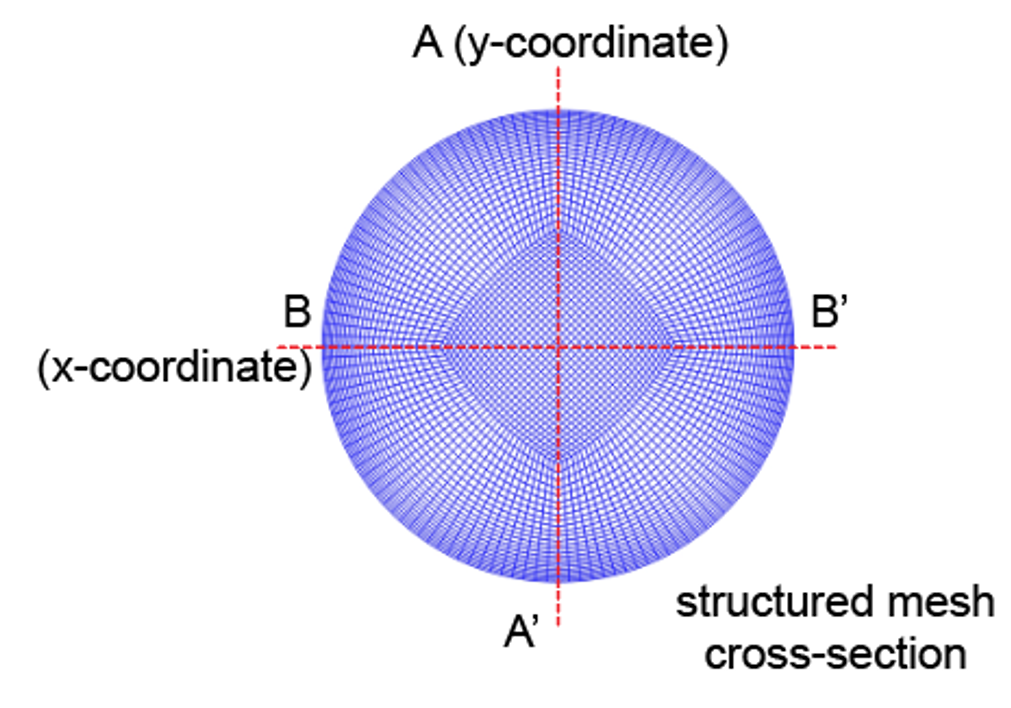}
		\caption{}
		\vspace*{4mm}
	\end{subfigure}
		~
	\begin{subfigure}[b]{0.7\textwidth}
		\includegraphics[width=\textwidth]{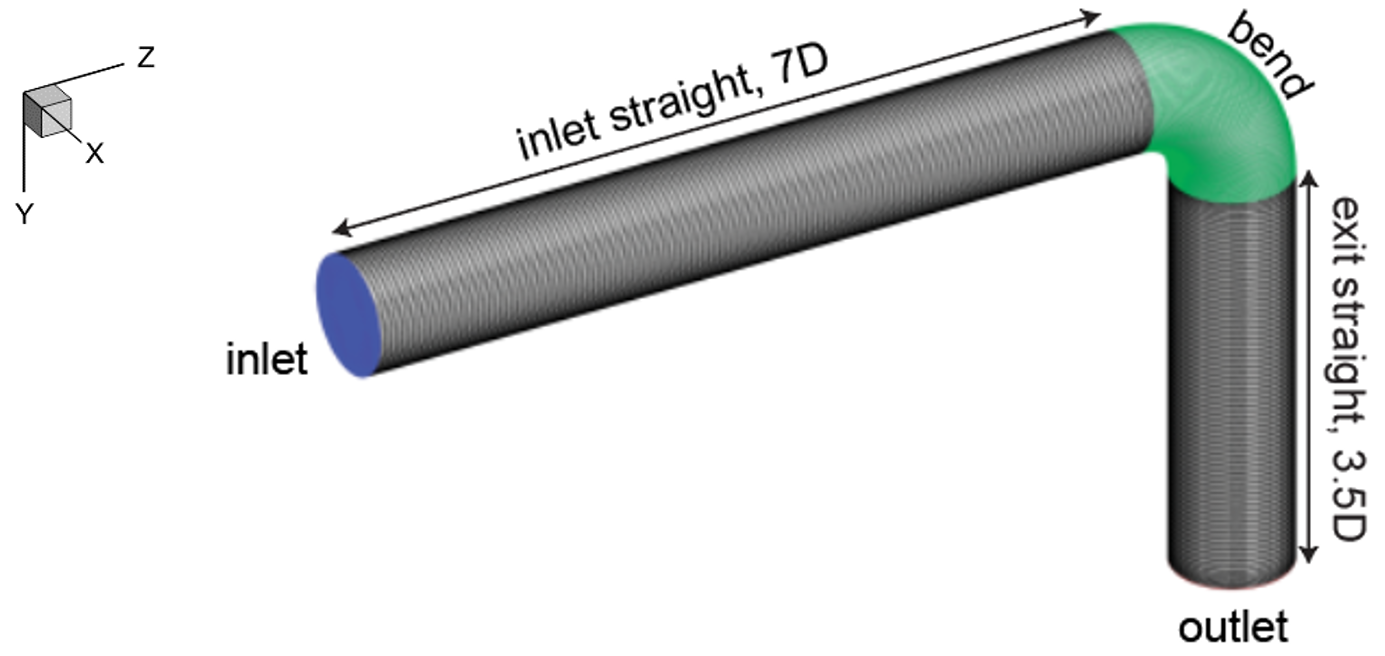}
		\caption{}
		\vspace*{4mm}
	\end{subfigure}
	\caption{(a) Schematic of the circular pipe bend geometry for all cases where $R_b$ and the pipe diameter, D varied. (b) Internal (cross-section) mesh which used O-grid block meshing, and concentration of cells near the wall. (c) Structured mesh on the external wall surface.}
\label{fig:geom}
\end{figure}

\clearpage
\begin{figure}
	\centering
	\includegraphics[width=0.7\linewidth]{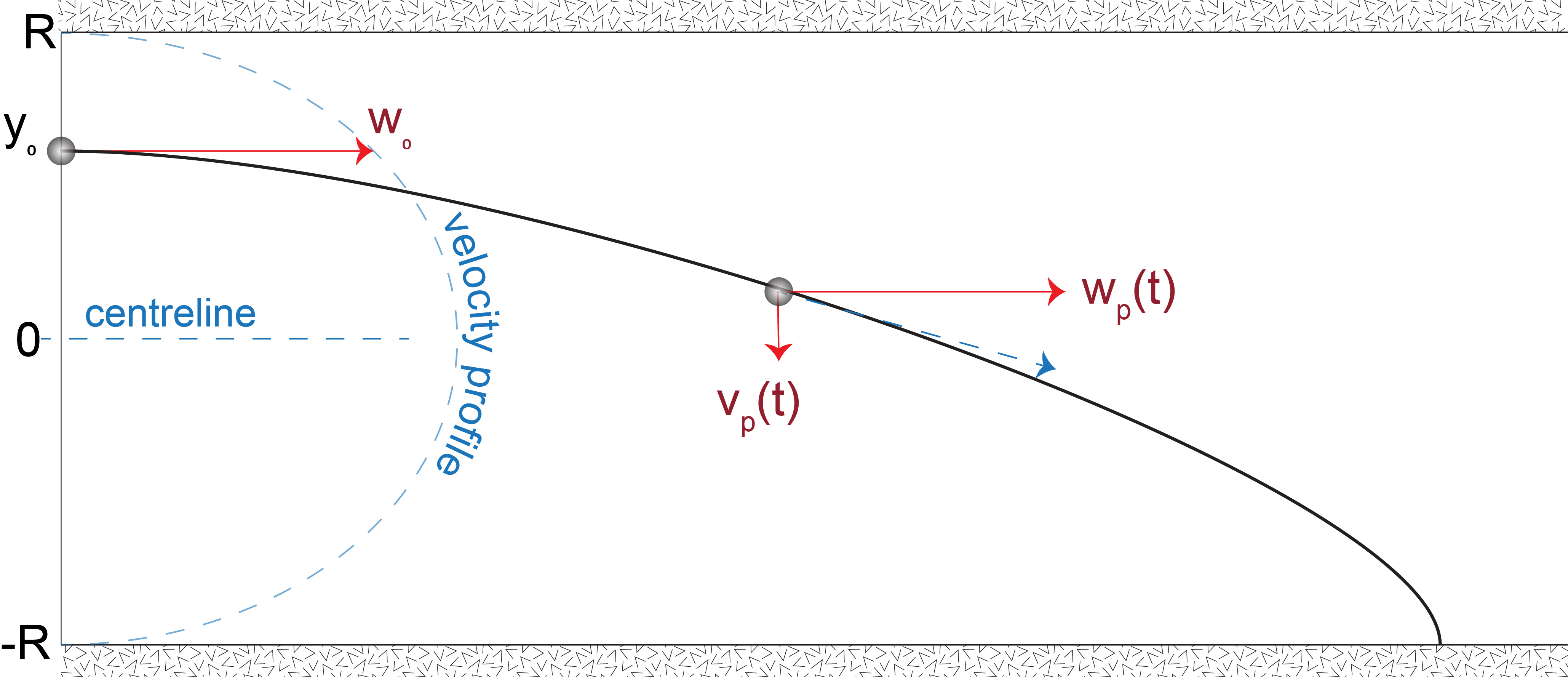}
	\caption{Schematic of the coordinate system for 2D particle motion for a fully developed velocity flow and particle momentum and sedimentation}
	\label{fig:sediment}
\end{figure}

\clearpage
\begin{figure}
	\centering
	\begin{subfigure}[b]{0.49\textwidth}
		\includegraphics[width=\textwidth]{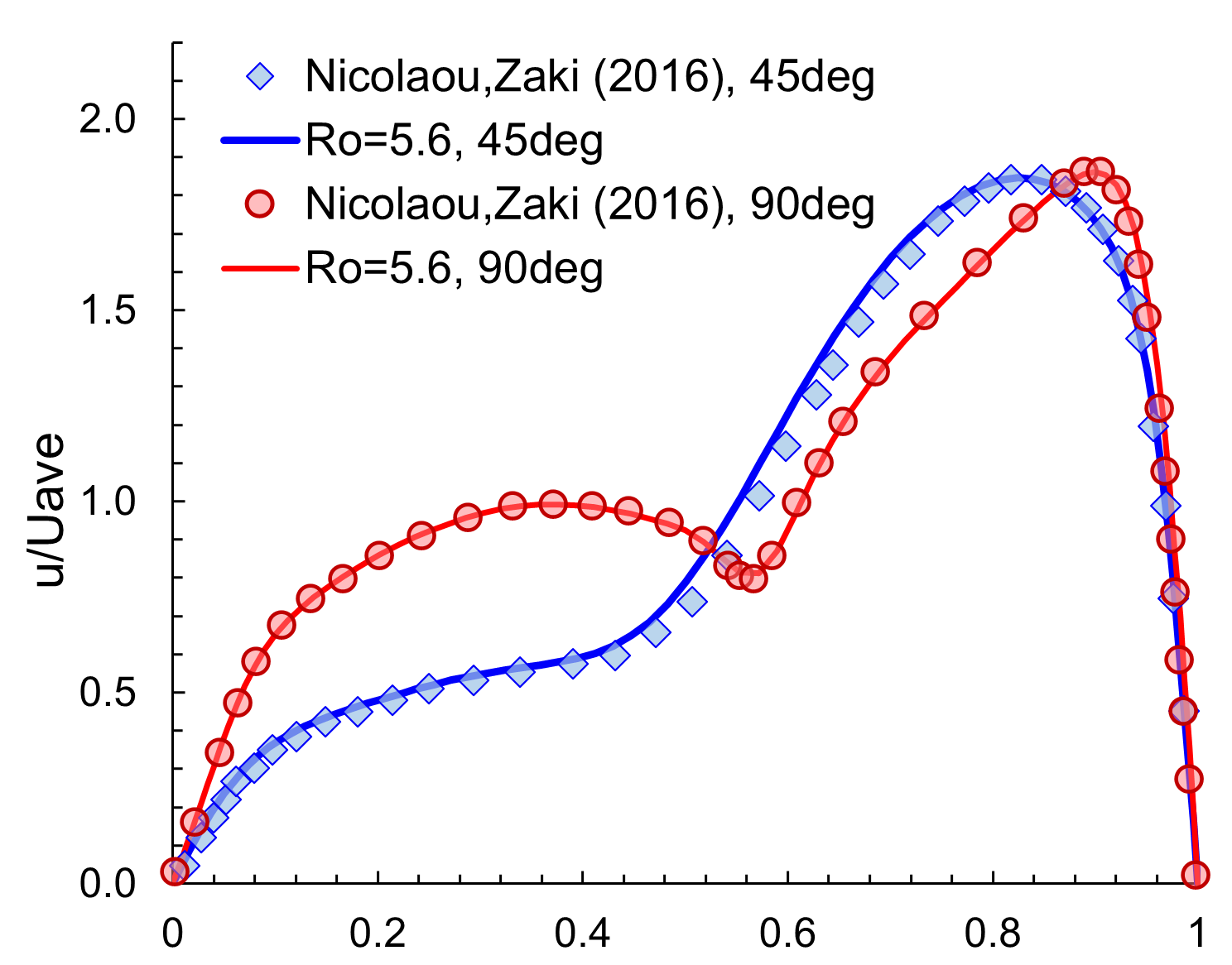}
		\caption{Line A-A' velocity profile}
		\vspace*{4mm}
	\end{subfigure}
	~
	\begin{subfigure}[b]{0.49\textwidth}
		\includegraphics[width=\textwidth]{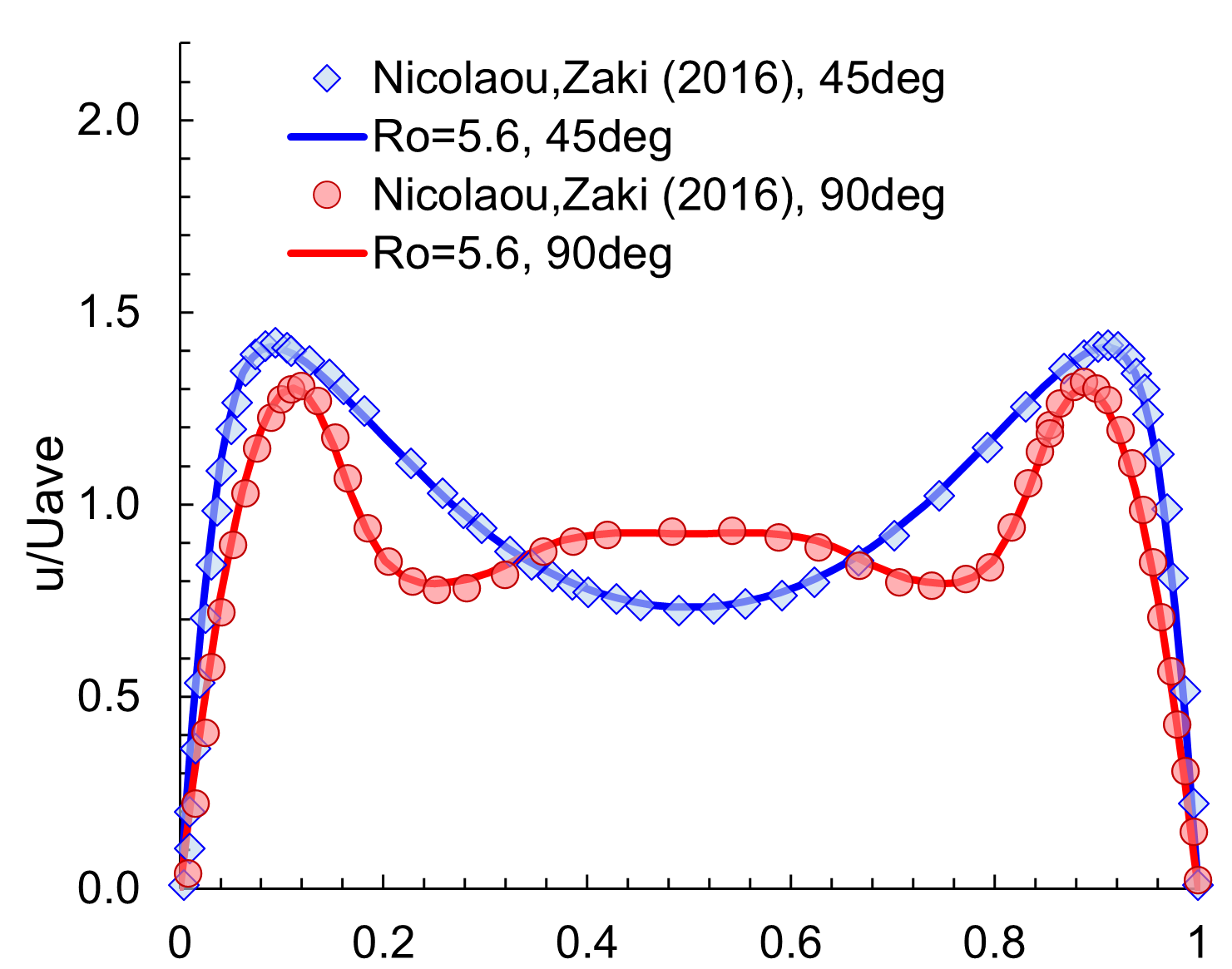}
		\caption{Line B-B' velocity profile}
		\vspace*{4mm}
	\end{subfigure}
	\caption{Radial velocity magnitude line profiles taken at $\theta = 45^\circ$ and $\theta = 90^\circ$ in the (a) x-coordinate (Line A-A'), and (b) y-coordinate (Line B-B')}
\label{fig:valid}
\end{figure}

\clearpage
\begin{figure}
	\centering
	\begin{subfigure}[b]{1\textwidth}
		\includegraphics[width=\textwidth]{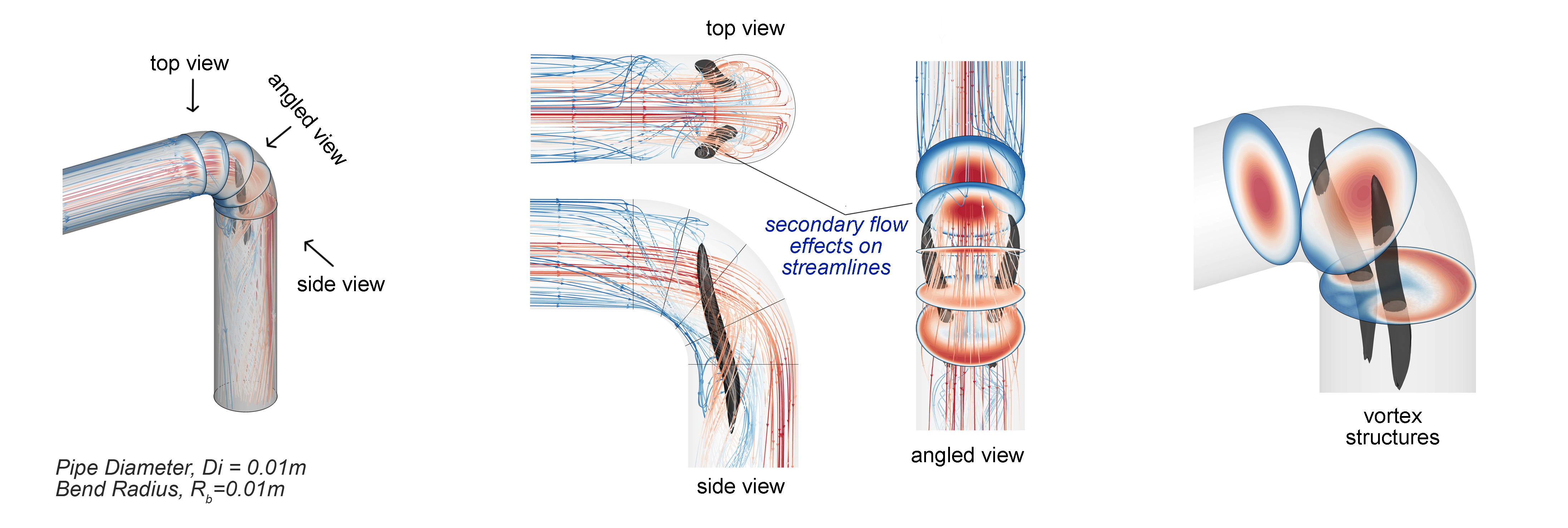}
		\caption{Pipe Diameter, $D_i = 0.01$m; Bend Radius, $R_b=0.01$m, $De = 707$}
		\vspace*{4mm}
	\end{subfigure}
	~
	\begin{subfigure}[b]{1\textwidth}
		\includegraphics[width=\textwidth]{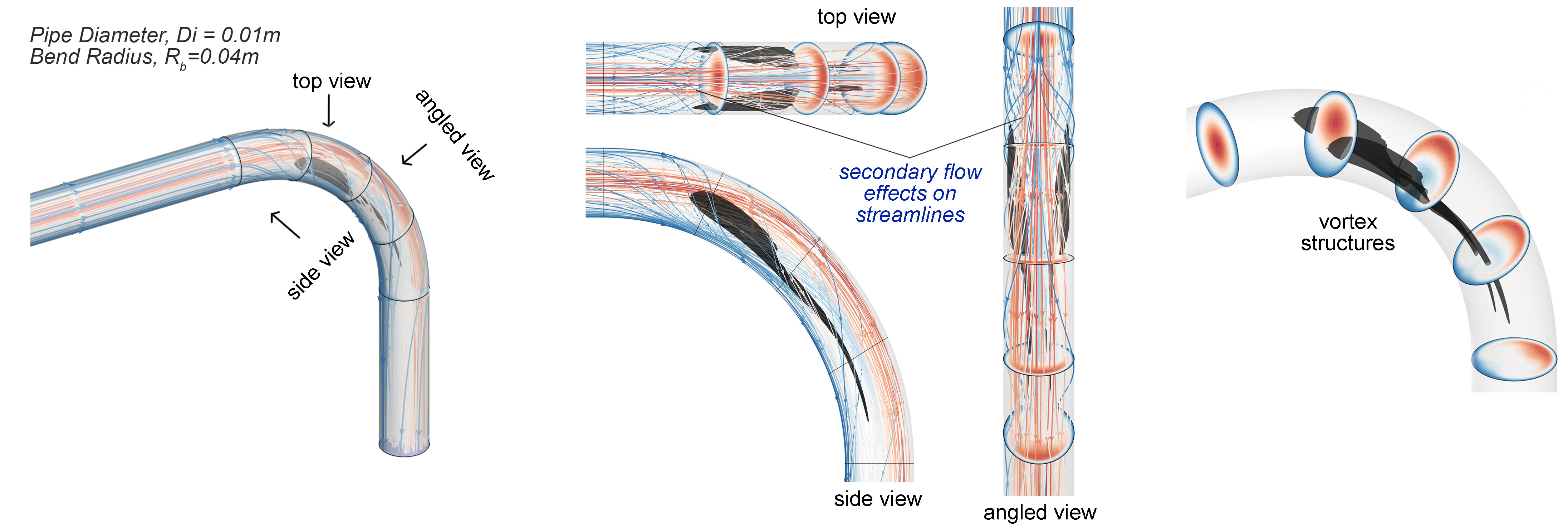}
		\caption{Pipe Diameter, $D_i = 0.01$m; Bend Radius, $R_b=0.04$m, $\mathrm{De} = 354$}
		\vspace*{4mm}
	\end{subfigure}
	\caption{Flow streamlines, contours, and vortex cores affected by radius of curvature for pipe diameter = 0.01 m, and Re=1000, for $R_o=2.0$ (sharp bend) and $R_o=8.0$ (longer bend)}
	\label{fig:streams}
\end{figure}

\clearpage
\begin{figure}
	\centering
	\includegraphics[width=1\linewidth]{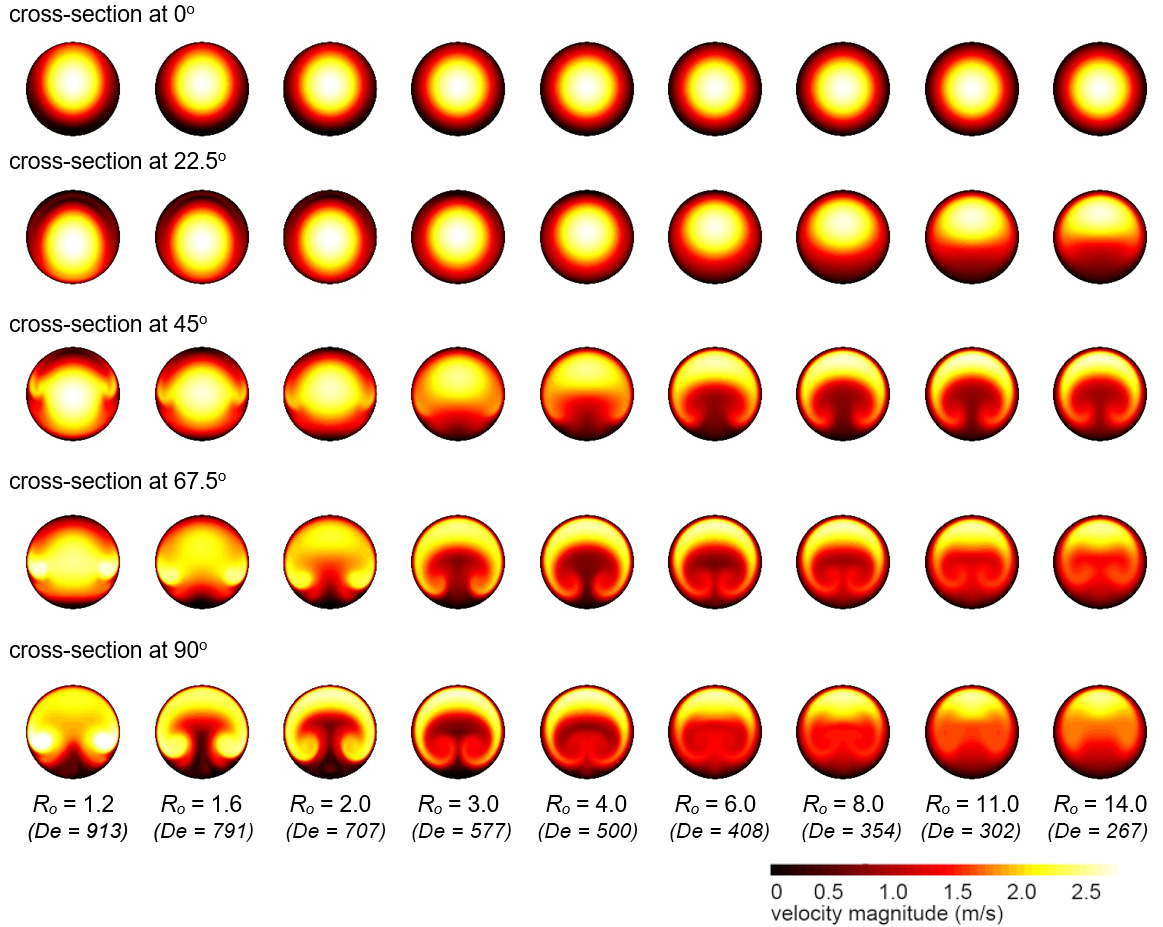}
	\caption{Schematic of the circular pipe bend geometry for all cases where Rb and the pipe diameter, D varied. (b) Pipe bend geometry comparison for Rb = 2.8D, and 1.0D. (c) Structured mesh on the external wall surface. (d) Internal (cross-section) mesh which used O-grid block meshing, and concentration of cells near the wall.}
	\label{fig:velCont}
\end{figure}

\clearpage
\begin{figure}
	\centering
	\begin{subfigure}[b]{0.49\textwidth}
		\includegraphics[width=\textwidth]{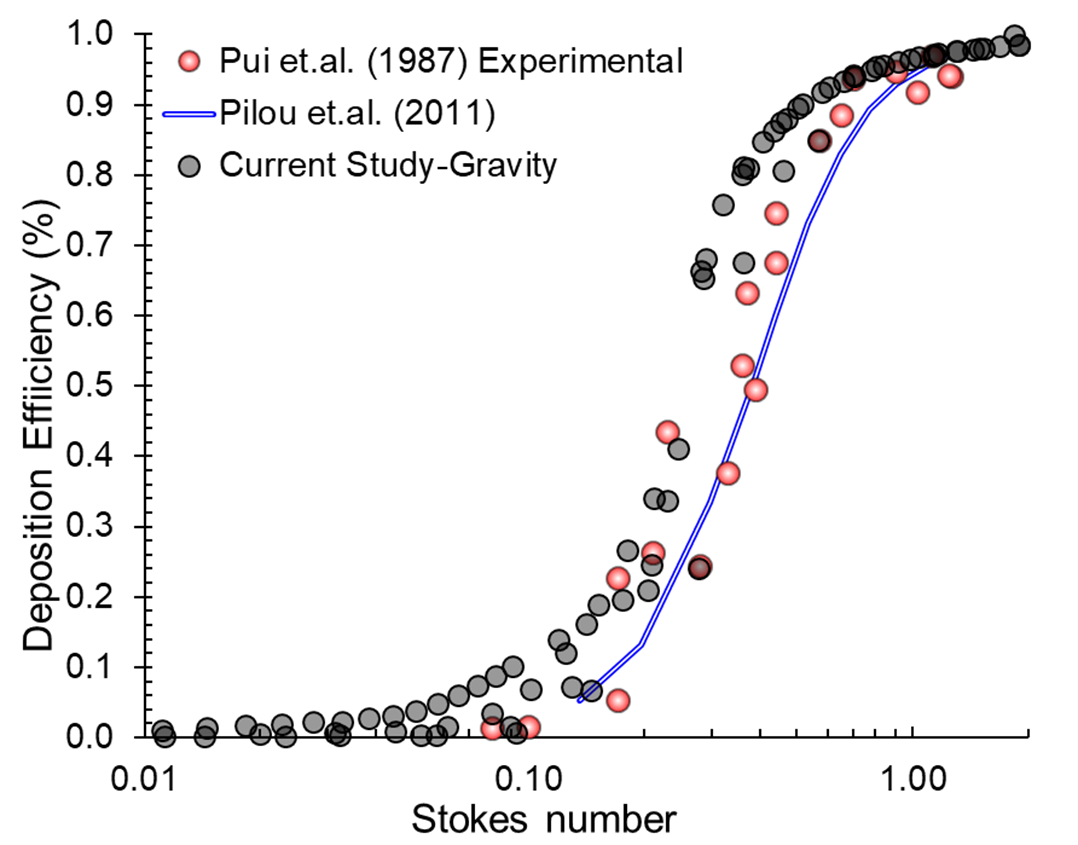}
		\caption{Deposition Efficiency with Gravity}
		\vspace*{4mm}
	\end{subfigure}
	~
	\begin{subfigure}[b]{0.49\textwidth}
		\includegraphics[width=\textwidth]{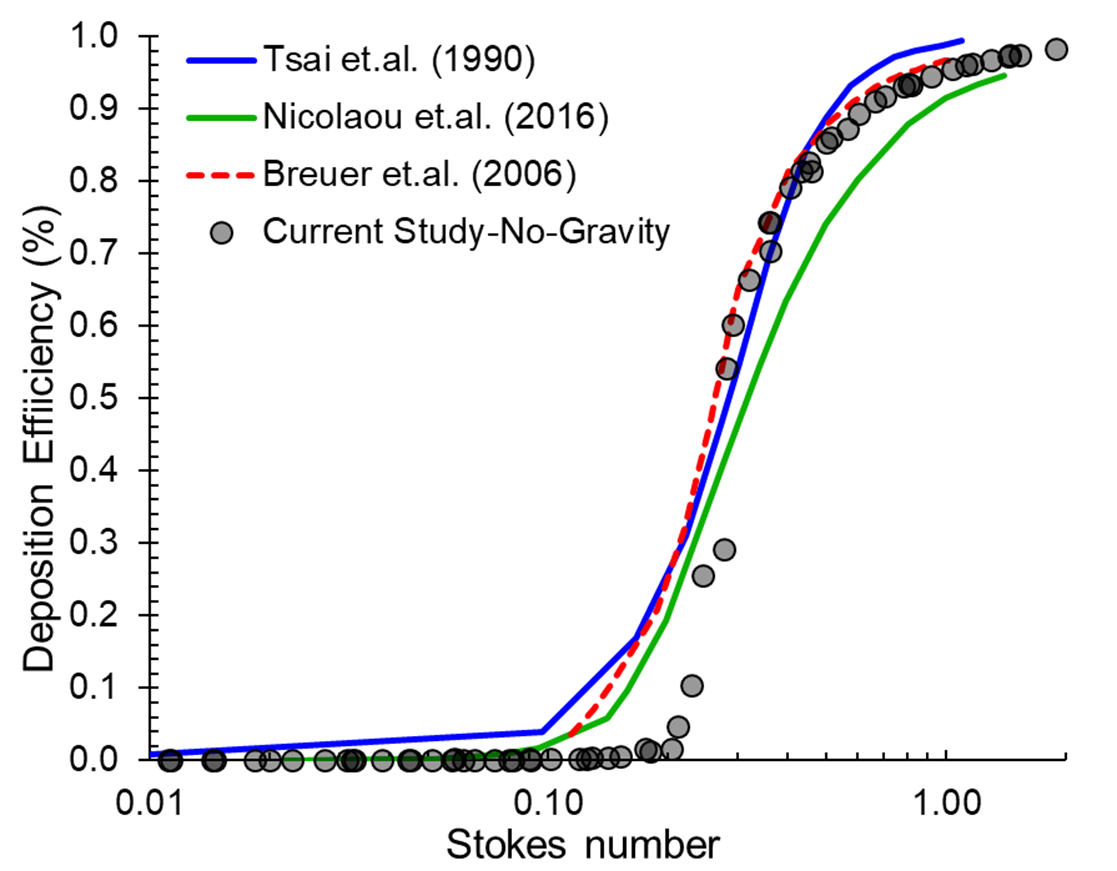}
		\caption{Deposition Efficiency without Gravity}
		\vspace*{4mm}
	\end{subfigure}
	\caption{Deposition efficiency profile for Re=1000, and $R_o = 5.6$, compared against studies in literature (a) with gravity, (b) without gravity.}
	\label{fig:depvalid}
\end{figure}

\clearpage
\begin{figure}
	\centering
	\includegraphics[width=1\linewidth]{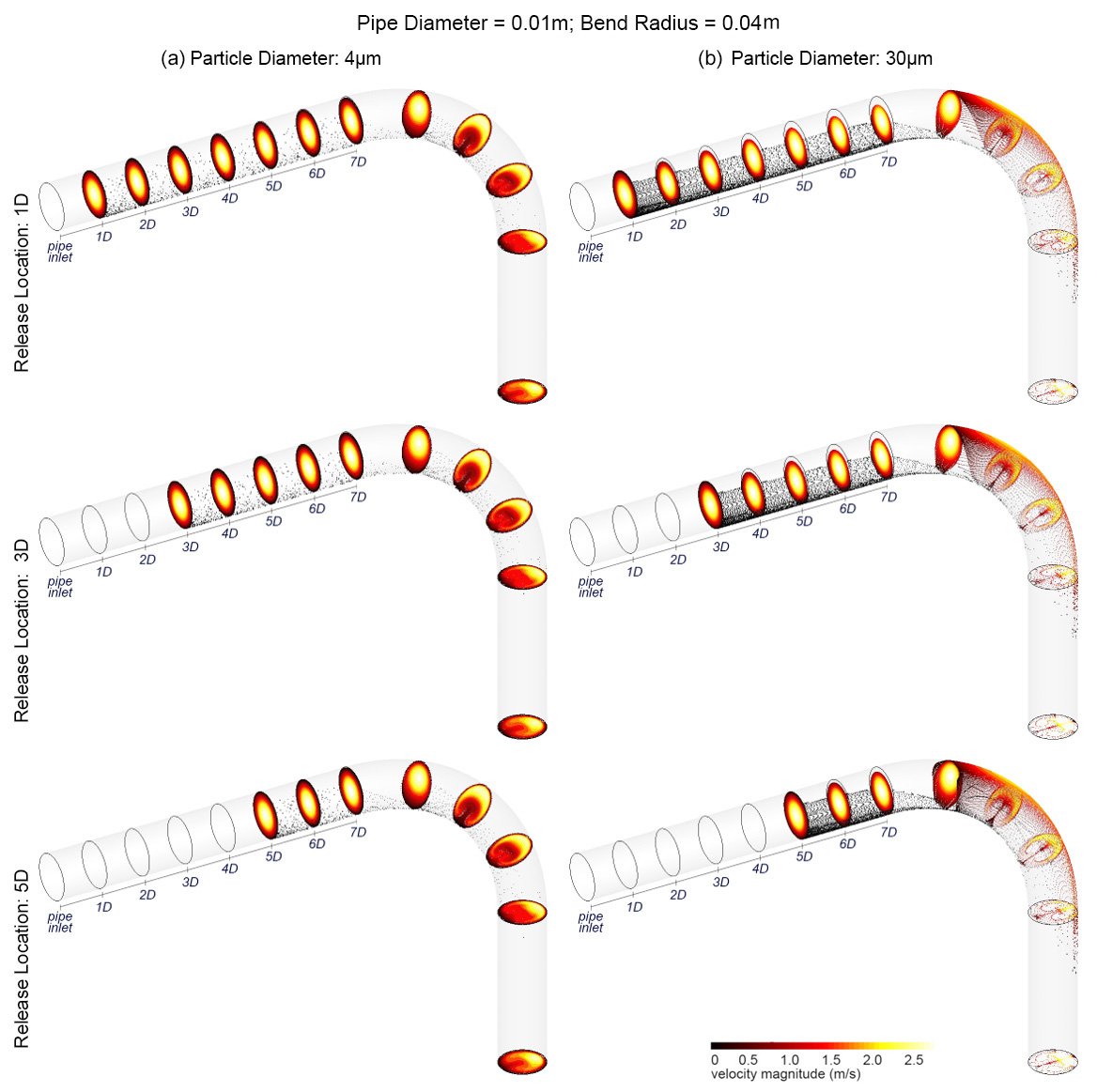}
	\caption{Deposition of $4\upmu$m and $30\upmu$m particles in a pipe with pipe diameter $D=0.01$m, bend radius $R_b = 0.04$m, and Re = 1000.}
	\label{fig:pdep11}
\end{figure}

\clearpage
\begin{figure}
	\centering
	\includegraphics[width=1\linewidth]{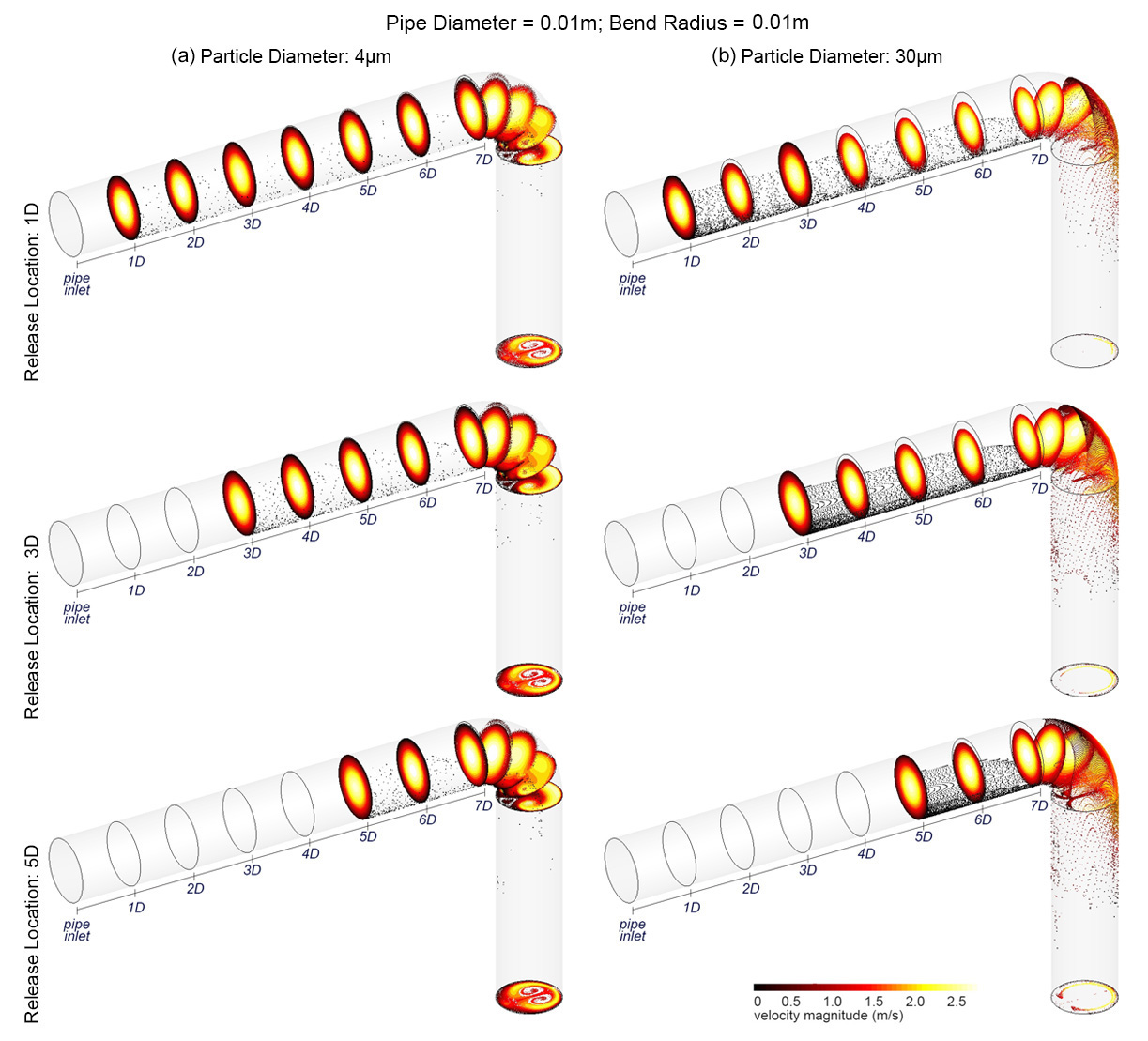}
	\caption{Deposition of $4\upmu$m and $30\upmu$m particles in a pipe with pipe diameter $D=0.01$m, bend radius $R_b = 0.01$m, and Re = 1000.}
	\label{fig:pdep14}
\end{figure}

\clearpage
\begin{figure}
	\centering
	\begin{subfigure}[b]{1\textwidth}
		\includegraphics[width=\textwidth]{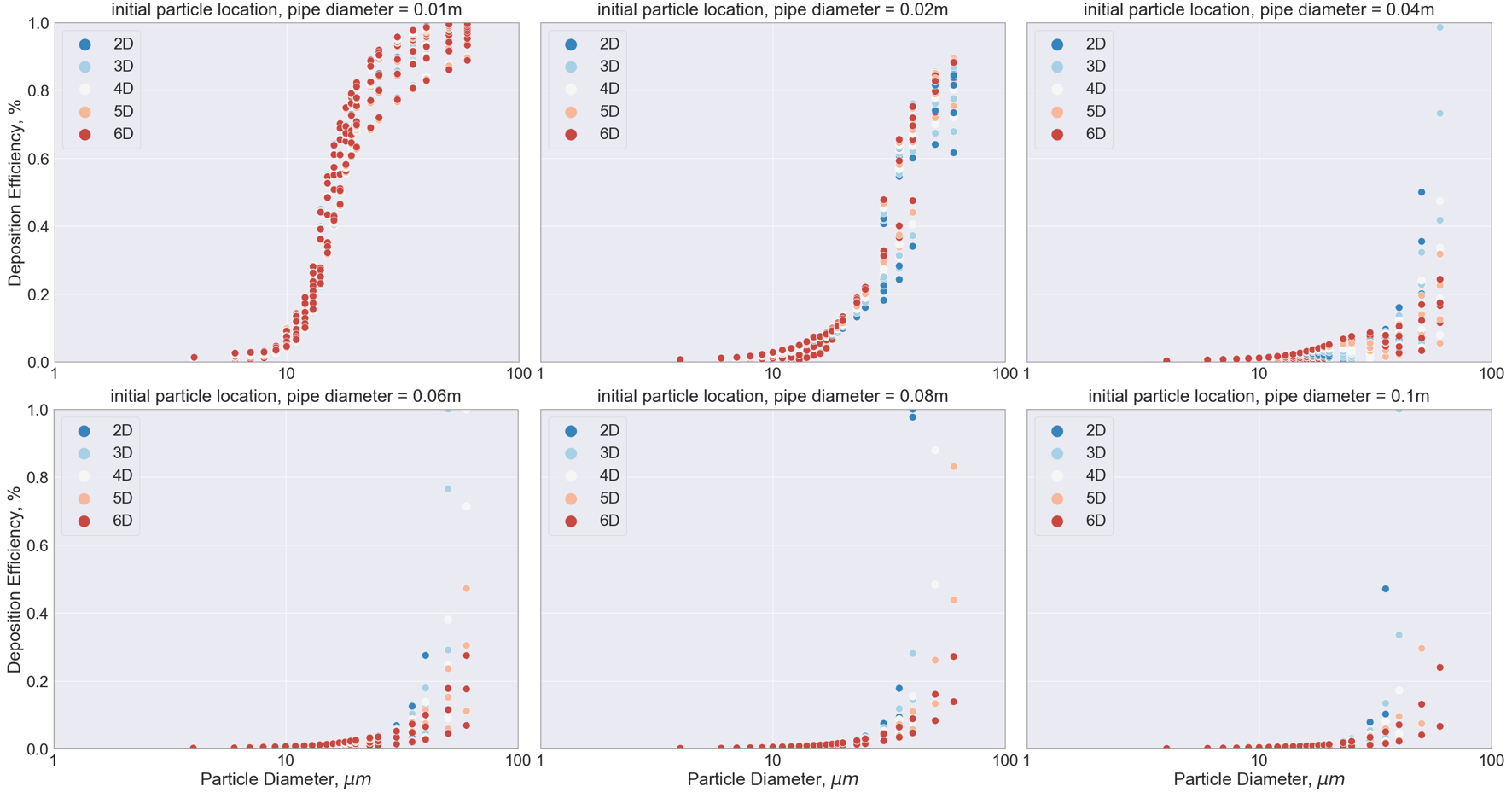}
		\caption{Deposition efficiency coloured by initial particle release location}
		\vspace*{4mm}
	\end{subfigure}
	~
	\begin{subfigure}[b]{1\textwidth}
		\includegraphics[width=\textwidth]{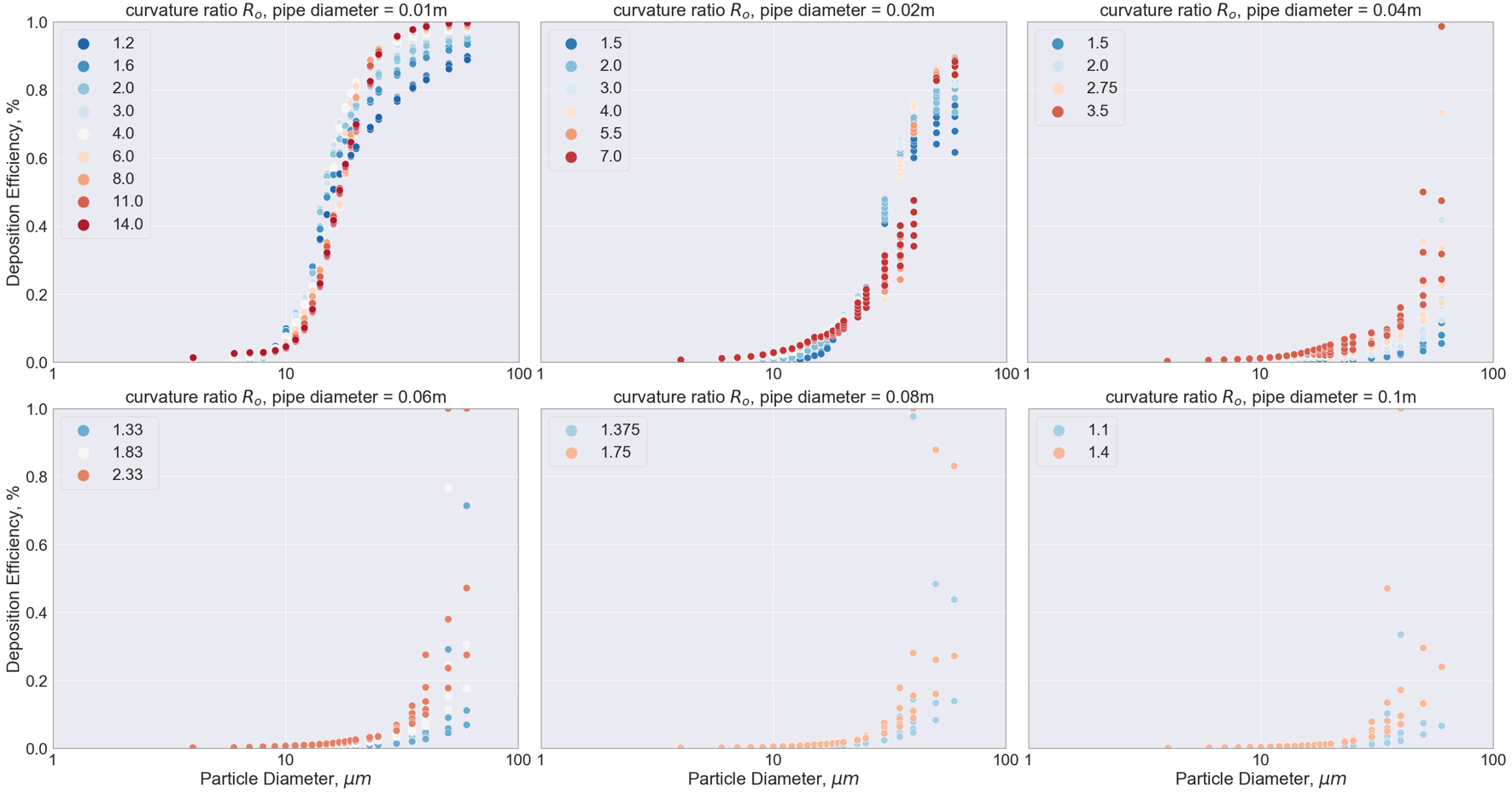}
		\caption{Deposition efficiency coloured by curvature ratio}
		\vspace*{4mm}
	\end{subfigure}
	\caption{Deposition efficiency on the pipe bend wall, due to (a) particle release location; and (b) bend curvature ratio}
	\label{fig:depcurves}
\end{figure}

\clearpage
\begin{figure}
	\begin{subfigure}[b]{1\textwidth}
		\centering
		\includegraphics[width=0.3\textwidth]{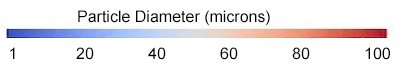}
	\end{subfigure}

	\begin{subfigure}[b]{0.49\textwidth}
		\includegraphics[width=\textwidth]{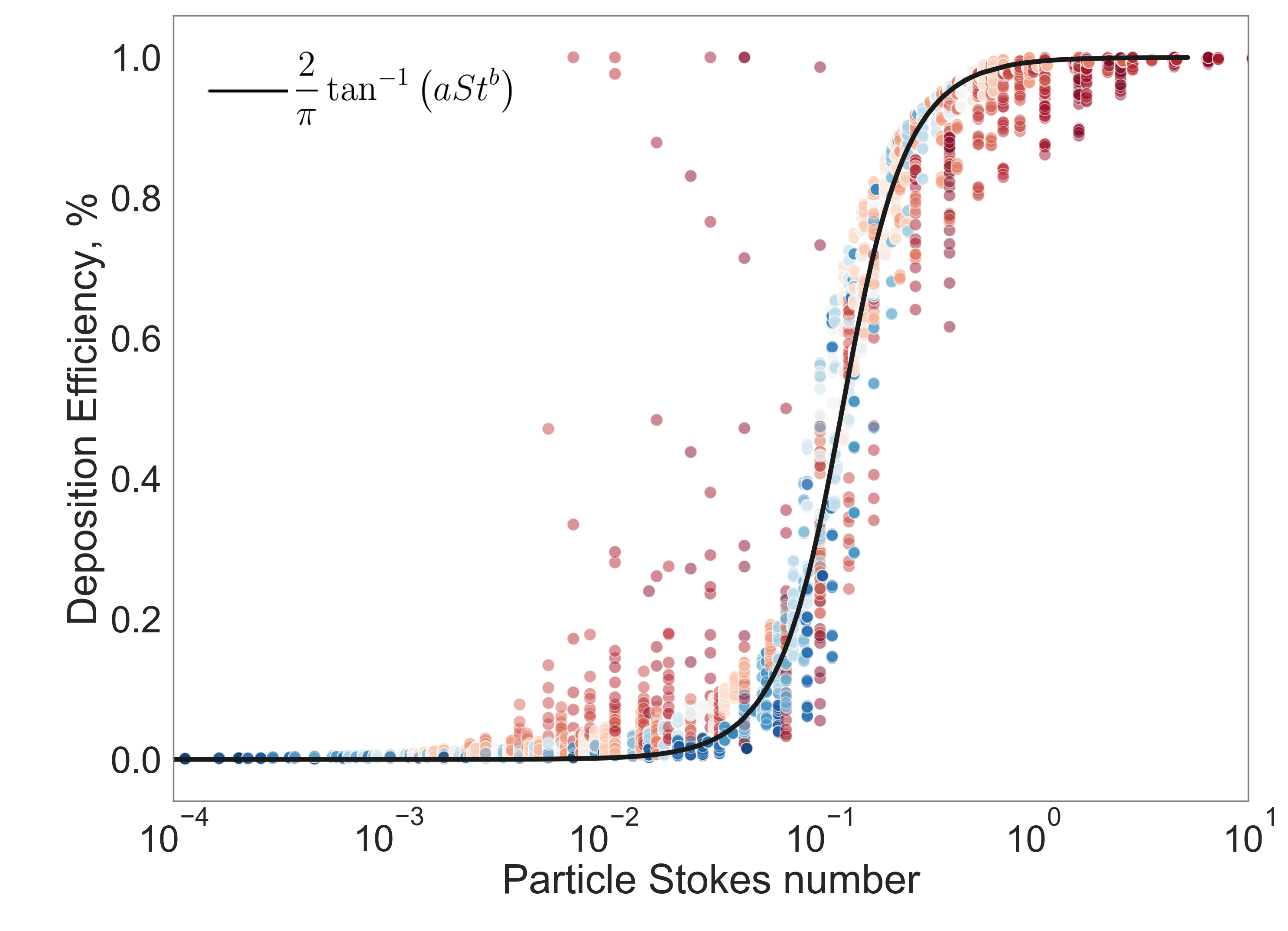}
		\caption{Deposition efficiency in the bend}
		\vspace*{4mm}
	\end{subfigure}
	~
	\begin{subfigure}[b]{0.49\textwidth}
		\includegraphics[width=0.95\textwidth]{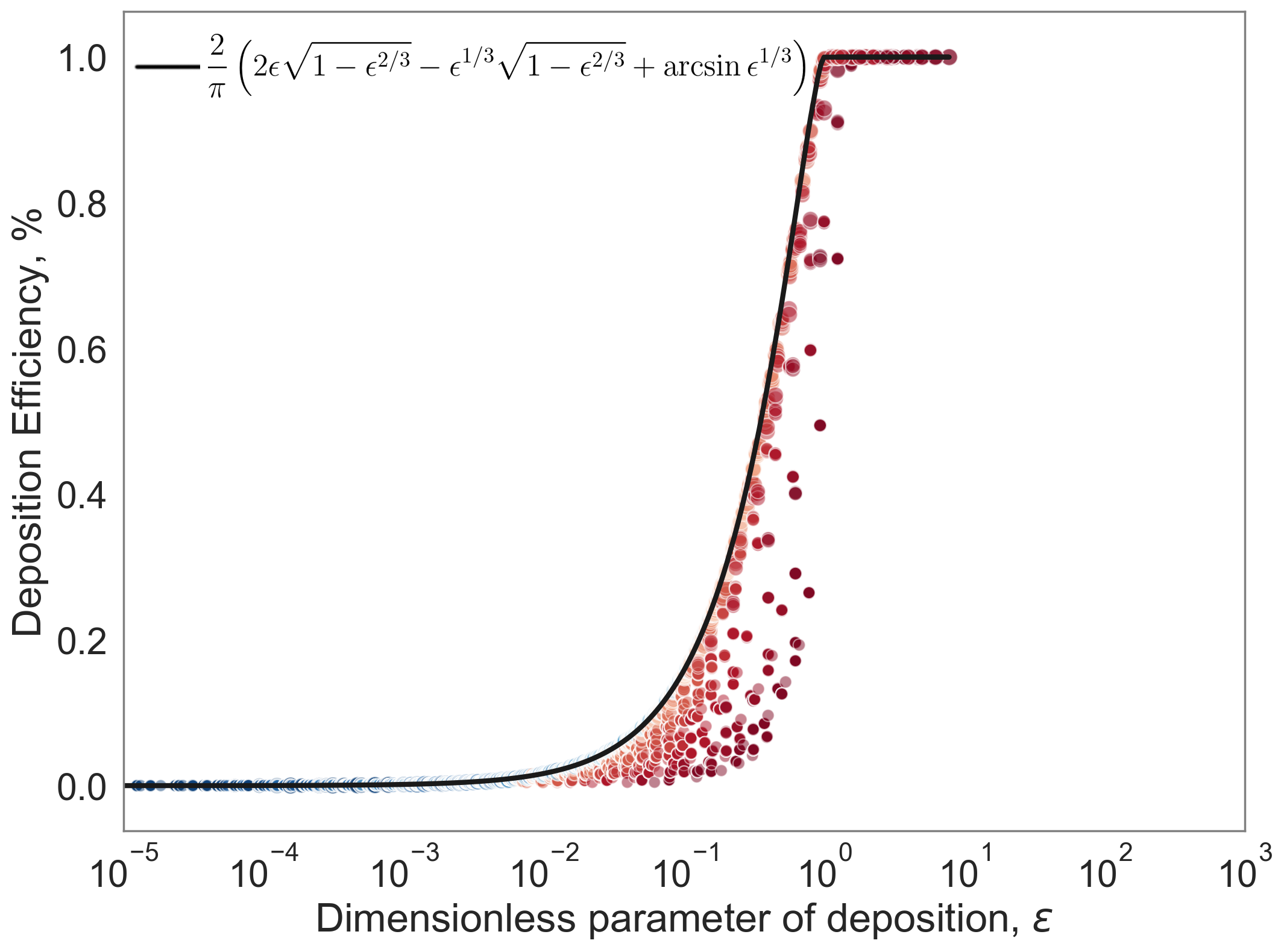}
		\caption{Sedimentation in the straight section}
		\vspace*{4mm}
	\end{subfigure}

	\caption{Deposition efficiency for all Dean numbers for (a) bend where $\mathrm{St} = \tau U_0 / r$ (\cite{Inthavong2019}); and (b) sedimentation in the straight section where $\epsilon= \frac{3}{4} \frac{V_s}{D} \frac{L}{U}$.  In both plots, the data points are semi-transparent and coloured by particle diameter, and therefore multiple data points that overlay on each other create a darker shade, while data points that are outliers are a lighter shade.}
\label{fig:correlations}
\end{figure}

\clearpage
\begin{figure}
	\centering
	\includegraphics[width=1\linewidth]{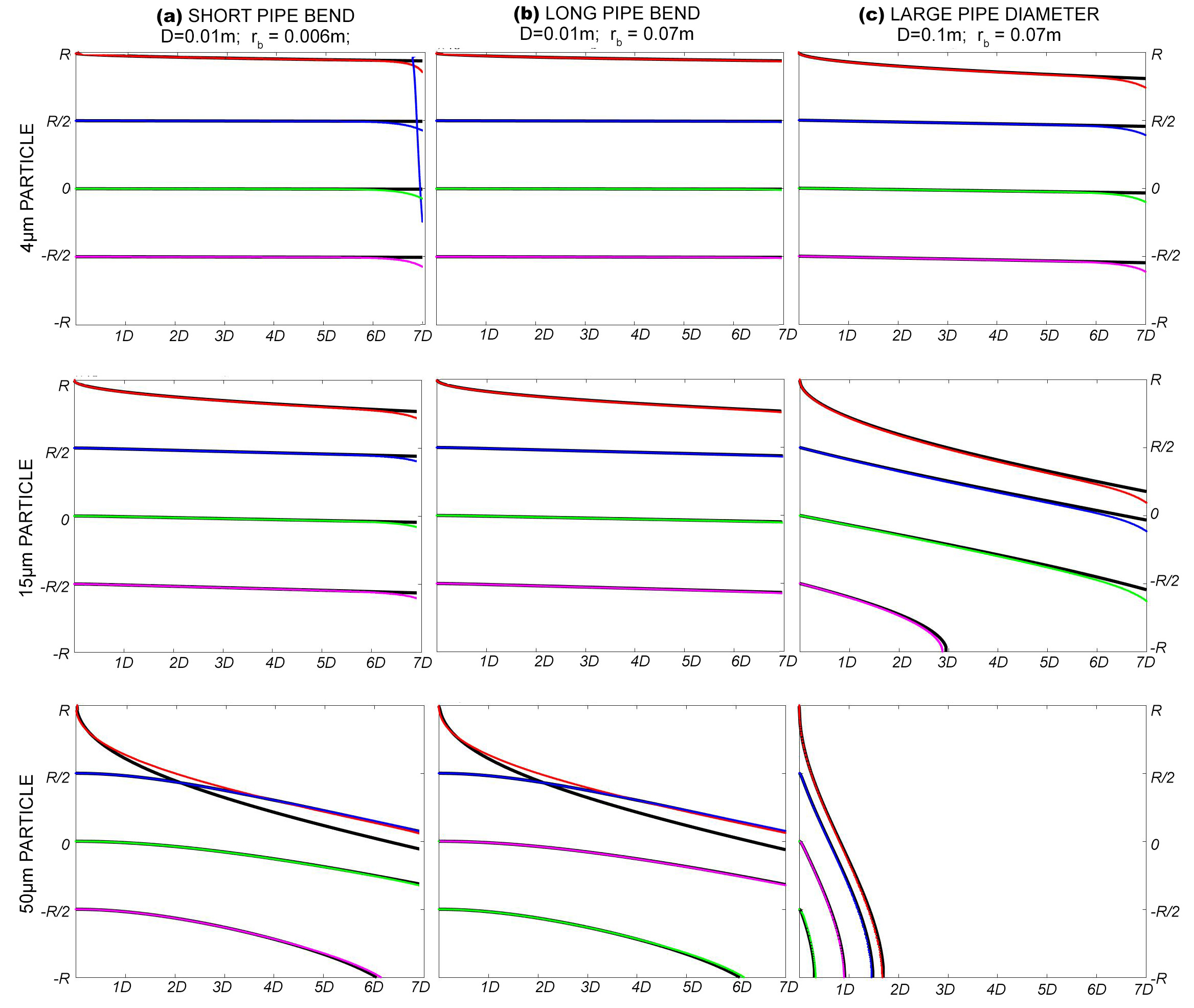}
	\caption{Comparison between analytical solution and simulation results (black lines) for particle trajectories in the straight pipe inlet section released from the inlet for (a) a short pipe bend with pipe diameter = 0.01m, and bend radius = 0.006m; (b) a long pipe bend with pipe diameter = 0.01m, and bend radius = 0.07m; and (c) a large pipe diameter with  pipe diameter = 0.1m, and bend radius = 0.07m.}
	\label{fig:sedimentTraj}
\end{figure}

\end{document}